\documentclass{caosp307}
\usepackage{graphicx}
\usepackage{epstopdf}
\usepackage{natbib}
\articleNo{XXX}
\pubyear{2021}
\volume{51}
\volnumber{2}
\firstpage{1}
\received{November 30, 2020}
\accepted{December 21, 2020}

\def\BibTeX{{\rm B\kern-.05em{\sc i\kern-.025em b}\kern-.08em
             T\kern-.1667em\lower.7ex\hbox{E}\kern-.125emX}}

\begin{document}
\hauthor{U.\,Munari et al.}
\title{Photometry and spectroscopy of the new symbiotic star 2SXPS J173508.4-292958}

\author{
        U.\,Munari \inst{1}   
      \and 
         P.\,Valisa \inst{2} 
      \and 
         A.\,Vagnozzi \inst{2} 
      \and 
         S.\,Dallaporta \inst{2}
      \and 
         F.~J.\,Hambsch \inst{2}
      \and 
         A.\,Frigo \inst{2}
        }
\institute{
          INAF Padova Astronomical Observatory, I-36012 Asiago (VI), Italy
          \and
          ANS Collaboration, Astronomical Observatory, 36012 Asiago (VI), Italy
          }
      
\date{XXX.X.XXXX}

\maketitle

\begin{abstract}
We present and discuss the results of our photometric and spectroscopic
monitoring of 2SXPS J173508.4-292958 carried out from April to August 2020. 
This X-ray source, in the foreground with respect to the Galactic center,
brightened in X-rays during 2020, prompting our follow-up optical
observations.  We found the star to contain a K4III giant with a modest but
highly variable H$\alpha$ emission, composed by a $\sim$470~km~s$^{-1}$ wide
component with superimposed a narrow absorption, offset by a positive
velocity with respect to the giant.  No orbital motion is detected for the
K4III, showing an heliocentric radial velocity stable at $-$12$\pm$1
~km~s$^{-1}$.  No flickering in excess of 0.005 mag~in~$B$ band was observed
at three separate visits of 2SXPS J173508.4-292958.  While photometrically
stable in 2016 through 2018, in 2019 the star developed a limited
photometric variability, that in 2020 took the form of a sinusoidal
modulation with a period of 38 days and an amplitude of 0.12 mag in $V$
band.  We argue this variability cannot be ascribed to Roche-lobe filling by
the K4III star.  No correlation is observed between the photometric
variability and the amount of emission in H$\alpha$, the latter probably
originating directly from the accretion disk around the accreting companion. 
While no emission from dust is detected at mid-IR wavelengths, an excess in
$U$-band is probably present and caused by direct emission from the
accretion disk.  We conclude that 2SXPS J173508.4-292958 is a new symbiotic
star of the accreting-only variety (AO-SySt).
\end{abstract}

\section{Introduction}

Attention to the X--ray source 2SXPS J173508.4-292958 has been recently
driven by a report from Heinke et al.  (2020) about a brightening of the
object they observed in X--rays with the Swift satellite in April 2020. 
Noting the positional coincidence with the cool 2MASS star
17350831-2929580 of Ks=7.4 mag, Heinke et al.  suggested that 2SXPS
J173508.4-292958 could be a previously unknown symbiotic star.  The optical
counterpart was quickly subjected to BVRI photometry and high- and
low-resolution spectroscopy by Munari et al.  (2020a) that found 2SXPS
J173508.4-292958 to host a K4 III/II giant, with marked near-UV excess and
H$\alpha$ in emission, the latter showing a broad profile with a central
absorption.  Such properties confirm a probable symbiotic nature for
2SXPS J173508.4-292958 (for a recent global review of symbiotic stars see
Munari 2019).  In this paper we report about the results of our
follow-up photometric and spectroscopic campaign to monitor 2SXPS
J173508.4-292958 (2SXPS for short in the rest of this paper) during the
period of seasonal visibility in 2020.

\section{Observations}

Our observations have been carried out primarily from Italy. At such
northern latitudes, the $-$29$^\circ$ declination of 2SXPS implies a very
short observing window, $\sim$2 hours around transit at meridian at
just 15 to 20$^\circ$ above horizon. Such a short observing interval coupled 
with the large atmospheric extinction and seeing degradation led to a
lower-than-usual S/N for our observations, and argued against employing the
highest possible resolution in the spectroscopic observations.

\subsection{$B$$V$$R$$I$ photometry}

We have obtained optical photometry of 2SXPS in the Landolt
photometric system from late April to early August 2020 with two telescopes
operated by ANS Collaboration.  ID 0310 is a 0.30m f/8 Richey-Chretien
telescope located in Cembra (TN, Italy) and equipped with an SBIG ST-8 CCD
camera, 1530$\times$1020 array, 9~$\mu$m pixels $\equiv$
0.77$^{\prime\prime}$/pix, with a field of view of
19$^\prime$$\times$13$^\prime$.  ID 1301 is a 0.50m f/6 Ritchey-Chretien
telescope located in Stroncone (TR, Italy).  It feeds light to an SBIG
STL1001E CCD Camera 1024$\times$1024 array, 24 $\mu$m pixels $\equiv$
1.60$^{\prime\prime}$/pix, field of view of 28$^\prime$$\times$28$^\prime$. 
Both telescopes adopt $B$$V$$R_{\rm C}$$I_{\rm C}$ photometric filters from
Astrodon, in the version corrected for red-leak.

	\begin{table}
	\small
	\begin{center}
		\caption{Our BVRI photometry on the Landolt system
                of 2SXPS J173508.4-292958.}
		\label{t1}
		\includegraphics[width=11.cm,clip=]{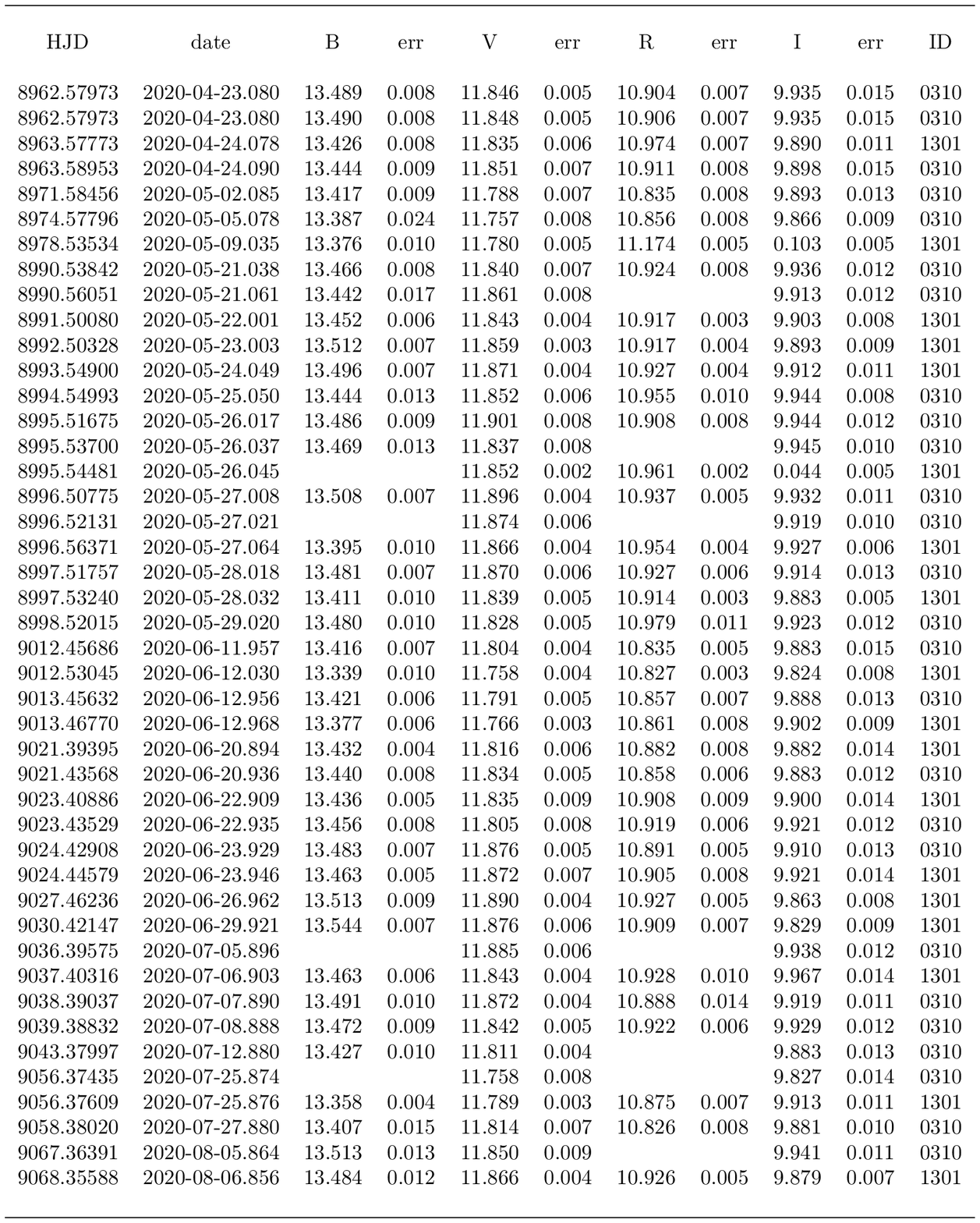}
	\end{center}
	\end{table}

Data reduction has involved all the usual steps for bias, dark and flat with
calibration images collected during the same observing nights.  We adopted
aperture photometry because the sparse field around 2SXPS did not required
PSF-fitting procedures.  The transformation from the local to the Landolt
standard system was carried out via color equations calibrated on a
photometric sequence recorded on the same frames as 2SXPS:
\begin{eqnarray}
\nonumber
V   &=& v + \alpha_v \times (v-i) + \gamma_v \\  \nonumber
B-V &=& \beta_{bv} \times (b-v) + \delta_{bv} \\ \nonumber
V-R &=& \beta_{vr} \times (v-r) + \delta_{vr} \\ \nonumber
V-I &=& \beta_{vi} \times (v-i) + \delta_{vi}    \nonumber
\end{eqnarray}
where lowercase and uppercase letters denote values in the local and
and in the standard system, respectively. The local photometric sequence
has been extracted from APASS DR8 survey (Henden et al.  2016), ported to
the Landolt system via the transformations calibrated by Munari et al. 
(2014).  Our photometry of 2SXPS is listed in Table~1.  The quoted errors
are the quadratic sum of the Poissonian error on the variable and the error
in the transformation to the standard system via the above color equations.

\subsection{Spectroscopy}

Low resolution spectra of 2SXPS have been obtained with a Shelyak LHIRES
spectrograph + 300 ln/mm grating mounted on the 0.50 Ritchey-Chretien
telescope operated for ANS Collaboration in Stroncone (TR, Italy). The 
CCD Camera is an ATIK~460EX (2749$\times$2199 pixels, 4.5~$\mu$m in size).
A slit width of 30~$\mu$m (=2.0 arcsec) has been adopted. Table~2 presents 
a log of these low-res observations.

Echelle high resolution spectra of 2SXPS have been obtained with the 0.84m
telescope operated by ANS Collaboration in Varese (Italy), and equipped with
a mark.III Multi-Mode Spectrograph from Astrolight Instruments.  The
detector is a SBIG ST10XME CCD camera (2192$\times$1472 array, 6.8 $\mu$m
pixel, KAF-3200ME chip with micro-lenses to boost the quantum efficiency). 
In the high resolution mode, an R2 Echelle grating of 79 ln/mm is used in
conjunction with an equilateral 60$^\circ$ prims made in high dispersion
N-SF11 flint glass, for a final 18,000 resolving power with a 2arcsec slit
width.  For 2SXPS observations a binning 2x2 was adopted leading to a
resolving power 10,000.

The spectroscopic observations at both telescopes were obtained in long-slit
mode, with the slit always rotated to the parallactic angle and widened to
2arcsec on the sky.  All data have been similarly reduced within IRAF,
carefully involving all steps connected with correction for bias, dark and
flat, sky subtraction, wavelength calibration and heliocentric correction. 
The low-res spectra have been flux calibrated using similar observations of
the nearby spectrophotometric standard HR 6698 observed immediately before
or after 2SXPS.  

	\begin{table}
	\small
	\begin{center}
		\caption{Log of spectroscopic observations
                of 2SXPS J173508.4-292958.}
		\label{t2}
		\includegraphics[width=9.5cm,clip=]{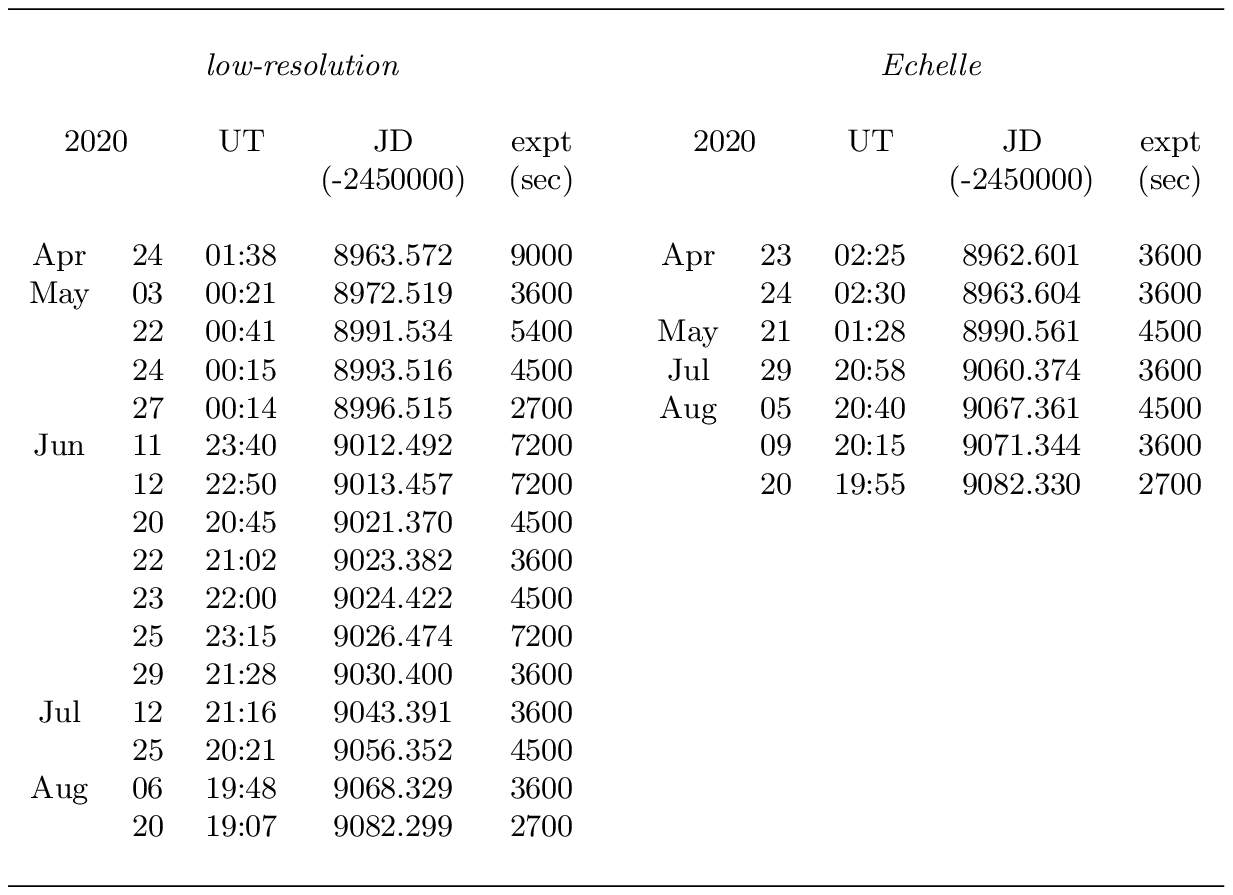}
	\end{center}
	\end{table}

\section{Results}

\subsection{Spectral classification, energy distribution, and radial
velocities}

	\begin{figure}
	\centerline{
		\includegraphics[width=11.cm,clip=]{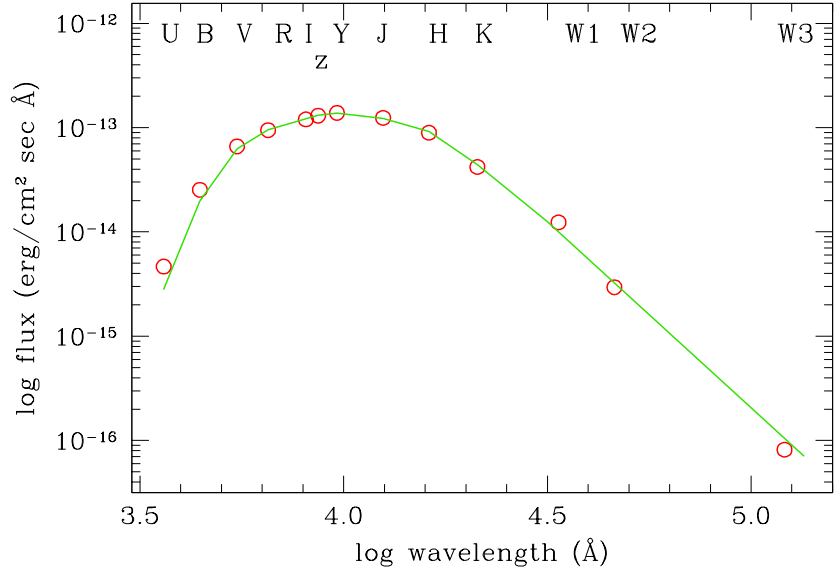}}
	\caption{The observed energy distribution of 2SXPS J173508.4-292958
        from our BVRIzY observations complemented with JHKs from 2MASS
        (Cutri et al.  2003), W1,W2,W3 from AllWISE (Cutri et al.  2014),
        and U band from Heinke et al.  (2020).  The green line is the
        intrinsic energy distribution of a K4III star from Koornneef (1983),
        Cousins (1980) and Fitzgerald (1970) subjected to an $E_{B-V}$=0.375
        following the reddening law of Fitzpatrick (1999).}
	\label{f1}
	\end{figure}

Even if projected close to the Galactic Center, 2SXPS is in the foreground
at 1.50 kpc distance according to Gaia eDR3 parallax 0.6685$\pm$0.0197 mas
(Gaia Collaboration 2020).  The distance would be minimally affected, to
1.53 kpc, by the application of the mean $-$0.017mas offset derived by
Lindegren et al.  (2020) to affect globally the eDR3 parallaxes (they
reported this bias to depend in a non-trivial way on - at least - the
magnitude, colour, and ecliptic latitude of the source, with different
dependencies that apply to the five- and six-parameter solutions in eDR3).

From Green et al.  (2019) 3D extinction maps, 2SXPS should suffer from a modest
E(B-V)=0.375.  We have dereddened accordingly our BVRI photometry combined
with 2MASS JHKs data and found an excellent match with the spectral energy
distribution of a K4III giant (Fig.~\ref{f1}).  Fitting instead with K3III or K5III
distributions would require to change E(B-V) to 0.49 or 0.28, respectively. 
Combining distance, reddening and observed V=11.825, an absolute magnitude
M(V)=$-$0.55 is derived, intermediate between K4III and K4III/II luminosity
classes for which Sowell et al.  (2007) list M(V)=$+$0.20 and M(V)=$-$1.00,
respectively.

The AllWISE mid-IR data in Fig.~\ref{f1} exclude the presence of circumstellar
dust warmer than $\sim$200~K. Any remnant blown-off by the progenitor of the
present-day degenerate companion to the K4III has therefore already dispersed 
at a great distance from the central binary.

The SED in Fig.~\ref{f1} is suggestive of an excess flux at the shortest
wavelengths, amounting to 0.29 mag at U band.  A word of caution is however
in order considering that the U-band {\it Swift} observation from
Heinke et al.  (2020) in Fig.~\ref{f1} is {\it not} contemporaneous with
the BVRIzY data, and 2SXPS is moderately variable (see below).  So the
apparent excess at the shortest wavelengths of Fig.~\ref{f1} may be (in part
or entirely) an artifact.  Such an ultraviolet excess is frequently
seen in symbiotic binaries of the accretion-only variety,  and probably
originates from the accretion disk around the companion to the cool giant
(Mukai et al.  2016, Munari et al.  2020b).  A significant contribution to
the UV excess can also be provided by nebular emission from ionized gas
located somewhere else in the binary system (Skopal 2005a,b), ionized by
either the hard radiation emitted from the accretion disk and its central
star or from some other mechanism, including wind collision.  For
simplicity, in this paper we attribute the UV excess to the accretion disk, 
including cumulatively in it also the contribution by any nebular region
distinct from the disk itself.

	\begin{figure}
	\centerline{
		\includegraphics[angle=270,width=11.cm,clip=]{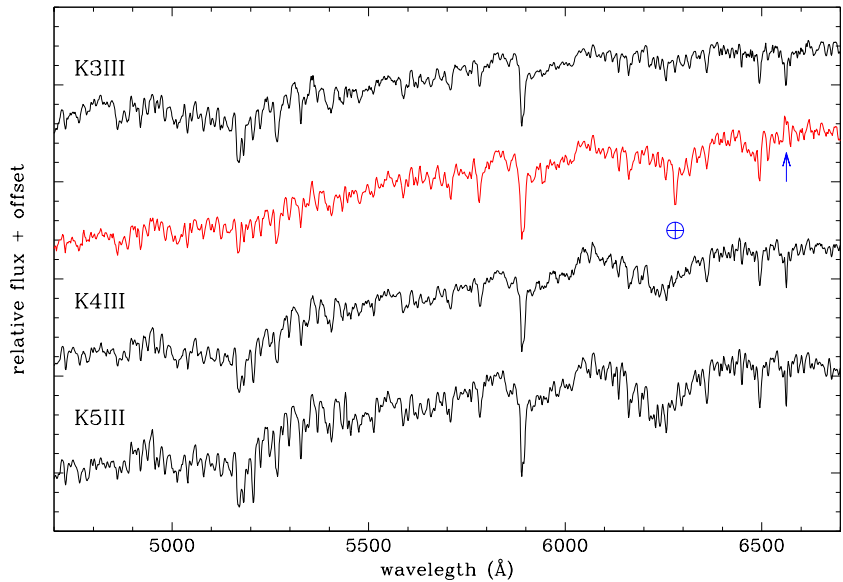}}
	\caption{The low resolution spectrum of 2SXPS J173508.4-292958 compared
        with those of K3III, K4III and K5III spectral standards observed
	along with it. The strong telluric O$_2$ band at 6285~\AA\ is
	marked, and the arrow points to H$\alpha$ in obvious emission.}
	\label{f2}
	\end{figure}

The K4III classification for 2SXPS agrees with our low resolution spectra. 
Their average is plotted in Fig.~\ref{f2} and compared with spectra of standards
for spectral types K3III, K4III and K5III selected from the compilation by
Yamashita and Nariai (1977) and observed along 2SXPS with the same
spectrograph set-up.  These nearby K giants trace the metallicity and
chemical partition of the Solar neighborhod.  The presence of H$\alpha$ in
emission in 2SXPS is rather obvious.  The very large airmass affecting all
our observations of 2SXPS marks the spectrum in Fig.~\ref{f2} with signatures of
telluric absorption red-ward of the NaI doublet at 5893~\AA, the strongest
feature being the O$_2$ band at 6285~\AA.  The BaII feature at 6496~\AA\
shows the same strength in 2SXPS and the field K giants of Fig.~\ref{f2},
suggesting a normal abundance (for a discussion of Barium enhanced stars vs. 
symbiotic stars see Jorissen 2003).

	\begin{table}
	\small
	\begin{center}
		\caption{Heliocentric radial velocities of the K-giant in
                2SXPS J173508.4-292958 obtained with the Varese 0.84m telescope.}
		\label{t2}
		\begin{tabular}{clcl}
			\hline
			&&\\
			\multicolumn{2}{c}{date} & RV$_\odot$ & err \\
                             && (km/s)     & (km/s) \\
			&&\\
                        2020 & Apr 23/24 &  $-$11  & 1.5 \\
                        2020 & May 21 &  $-$15  & 3   \\
                        2020 & Aug 05 &  $-$13  & 1.5 \\
                        2020 & Aug 09 &  $-$12  & 1.5 \\
                        2020 & Aug 20 &  $-$12  & 2   \\
			&&\\
			\hline
		\end{tabular}
	\end{center}
	\end{table}

Radial velocities have been obtained from the Echelle spectra by
cross-correlation (fxcor task in IRAF) with spectra of the IAU radial
velocity standard HR 6859 obtained immediately before and after those of
2SXPS on each visit.  This standard is rather convenient: it is located a
short angular distance from 2SXPS at RA=18:21:48 DEC=$-$29:49:18 (minimizing
differential spectrograph flexures), it is very bright at V=2.7 mag, and its
K2III spectral type is an excellent proxy for that of 2SXPS.  We adopted the
$-$20.0$\pm$0.0~km~s$^{-1}$ heliocentric velocity listed in the "Standard
Radial Velocity Stars" section of the Nautical Almanac.  The epoch radial
velocities so obtained for 2SXPS are listed in Table~3.  They are fully
compatible with a constant heliocentric radial velocity of $-$12$\pm$1
~km~s$^{-1}$ for 2SXPS during the four months span of our observations. 
This suggests that the orbital period is either rather long (and as a
consequence the amplitude of orbital motion quite small) and/or the viewing
angle to 2SXPS is oriented high above the orbital plane.  The
tangential velocity of 2SXPS is similarly low at $-$26~km~s$^{-1}$ (from
Gaia eDR3 parallax and proper motions), for a combined space velocity of
29~km~s$^{-1}$.
 
\subsection{The variable and structured emission in H$\alpha$}

Supporting the association of the K4III giant with the X--ray source 2SXPS
is the presence on optical spectra of emission in H$\alpha$, not expected in
normal and single K4III stars belonging to the field. 

\begin{figure}
	\centerline{
		\includegraphics[width=12.cm,clip=]{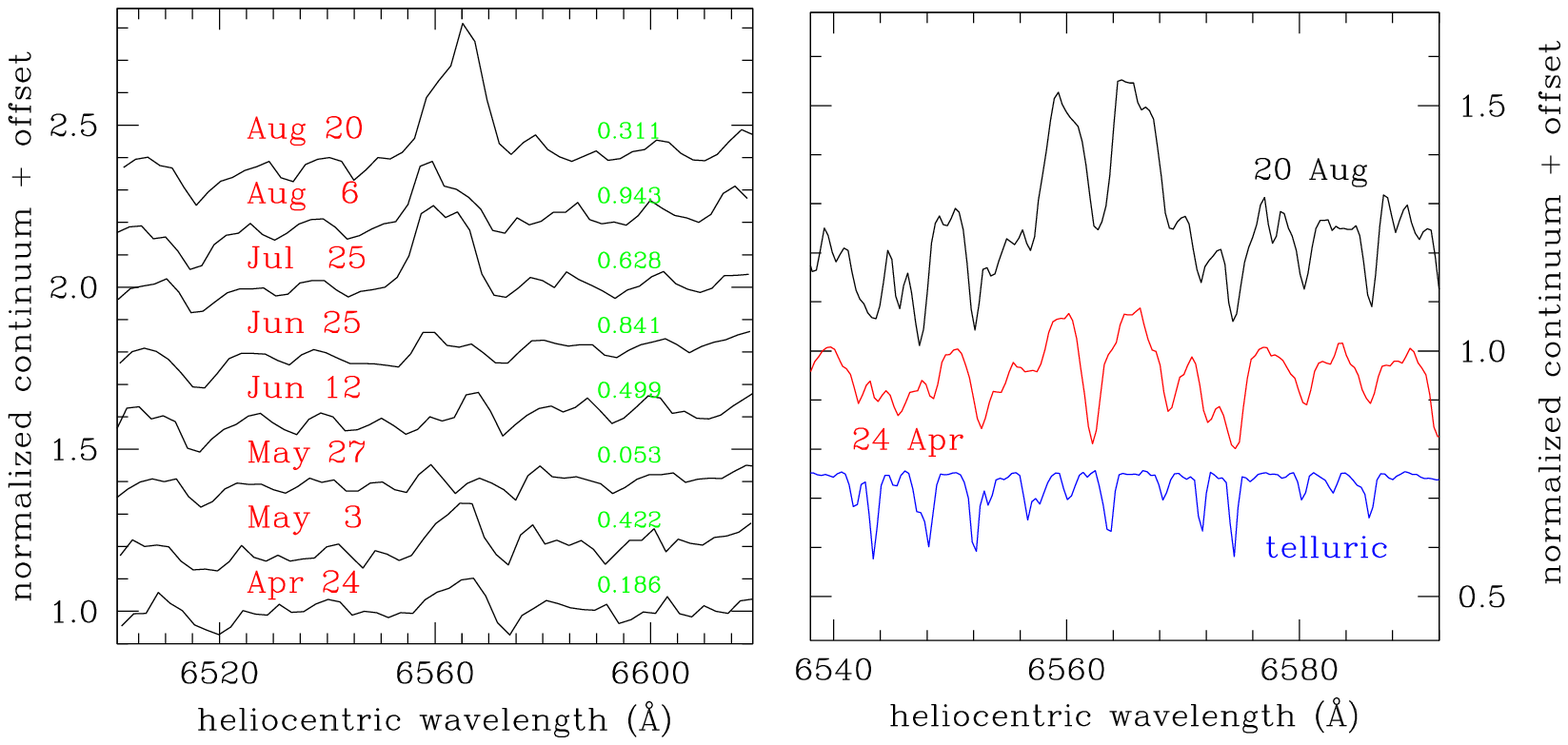}}
	\caption{The variable H$\alpha$ emission observed in 2SXPS
        J173508.4-292958 during 2020.  {\it Left panel}: a strong variation
        in intensity in recorded on our low-res spectra.  {\it Right panel}:
        on Echelle spectra the profile turns out to be characterized by a
        broad emission component with superimposed a narrow absorption.
        For reference, the spectrum at the bottom show the telluric
	absorption lines in the region as derived from observations of
        a telluric divider observed along the program star.}
	\label{f3}
	\end{figure}

In the left panel of Fig.~\ref{f3} we present small portions around H$\alpha$
from a sample of our low-resolution spectra (those with the highest S/N). 
The flux and equivalent width of H$\alpha$ are observed to change by a large
amount over the course of our observations; no correlation is noted (see
phases marked on the spectra) with the low-amplitude (0.12 mag), 38-day
sinusoidal variability exhibited by 2SXPS, and presumably associated with
the K4III giant (see sect.  3.4 an Eq.  1 below).  The absence of
correlation, and the low temperature characterizing a K4III giant, suggest
the accretion disk around the companion as the origin of the H$\alpha$
emission.

The H$\alpha$ profile as recorded on our Echelle spectra is presented on the
right panel of Fig.~\ref{f3}. It is composed by a broad emission, about 470
km~s$^{-1}$ of width at half intensity, on which it is superimposed a narrow
absorption component (only slightly wider than the instrumental PSF), which
heliocentric velocity averaged over all spectra is $-$4~km~s$^{-1}$. While
the observed velocity of the K4III giant is consistent with a constant
value, the velocity of the narrow absorption component could vary over the
range $-$10 to 0 ~km~s$^{-1}$. Our spectra lack however the S/N required for
a firm conclusion about that.

One thing about the narrow absorption seems however well established: its
velocity is {\it positive} with respect to the K4III giant, so it seems
unlikely it may form in the gentle wind out-flowing from it and engulfing
the whole binary system.  A typical shift $\Delta$vel $\approx$$-$10/$-$20
~km~s$^{-1}$ is generally observed in symbiotic stars for the narrow
absorption superimposed to the H$\alpha$ emission, matching the typical wind
velocity of field cool giants (Munari et al.  2020b,  Shagatova et al. 
2020).  This leaves open the possibility that the narrow absorption
component seen in 2SXPS may form elsewhere in the binary system (as the
accretion disk itself or an atmosphere engulfing it), or in alternative its
appearance is spurious and the H$\alpha$ profile is actually composed by two
separate emission components.  The latter could be expected for an accretion
disk seen at high inclination angle (e.g.  Horne and Marsh 1986), but this
condition seems to contrast with the limited (or null) orbital motion
exhibited by the K4III giant over the four months of our observations.

\subsection{No flickering observed}

We carried out a search for flickering at optical wavelengths as an
indication of on-going accretion.  To search for flickering we used a 50cm
telescope, with a 40arcmin corrected field of view and quality
photometric filters, operated robotically for ANS Collaboration in Atacama
(San Pedro Martir, Chile).  We observed in $B$-band with 1-min integration
time, with interspersed observations in $V$-band serving to construct the
$B$$-$$V$ color base for the transformation from the instantaneous local
photometric system to the Landolt's standard one.  The same local
photometric sequence and data reduction procedures as used for the long term
$B$$V$$R$$I$ monitoring, were adopted.

	\begin{figure}
	\centerline{
		\includegraphics[width=12.cm,clip=]{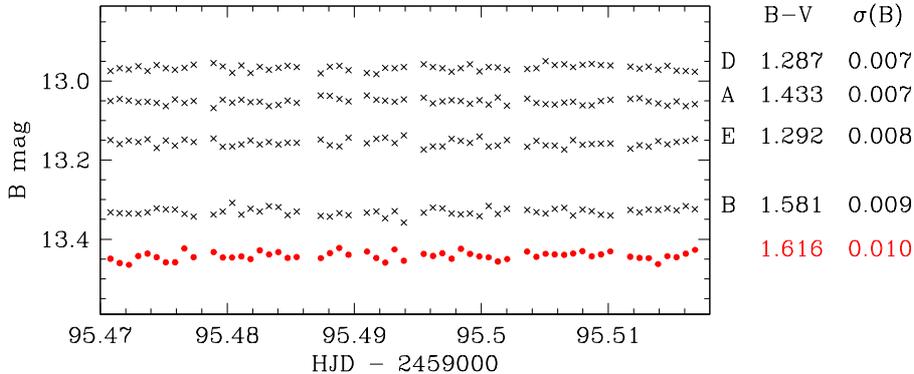}}
	\caption{Results from a 70min run in $B$ band in search for flickering
        from 2SXPS J173508.4-292958 (red dots). The regular gaps in the data
        mark epochs when $V$-band exposures have been obtained to allow
        the transformation to the standard system. The corresponding data
	for four field stars of similar color and magnitude are presented as
        crosses. The progression from $\sigma$=0.007 for field star D to
        $\sigma$=0.010 for 2SXPS is accounted for by the $\Delta
	B$$\sim$0.5mag difference, and exclude the presence of any
	flickering with an amplitude in excess of 0.005 mag.}
	\label{f4}
	\end{figure}

About 50 field stars close on the image to 2SXPS, of a similar magnitude and
well isolated from neighboring stars were also measured on all recorded
images in exactly the same way as 2SXPS.  The photometry of these 50 field
stars was then inspected looking for those with a $B$$-$$V$ color as close
as possible to 2SXPS.  Four such stars were found, which serve as samplers
of the observational noise above which the flickering has to be detected.

A first run, 70min in duration, was carried out from Sept 2.971 to 3.018 UT,
2020.  The results are presented in Fig.~\ref{f4}: the dispersion of $B$-band
measurements for 2SXPS is the same as for the four field stars of the same
color and magnitude, and if any flickering is present, its amplitude does
not exceed 0.005 mag.  Other two runs, each one lasting 35min were carried
out from Oct 24.980 to 25.003, and from Oct 26.981 to 27.003 UT, 2020.  They
confirmed the findings for the early September run, with no flickering being
detected.

We have seen above how large is the variability displayed by H$\alpha$
emission. It probably parallels a similarly large variability of the accretion
rate onto the compact object in 2SXPS. Therefore, the absence of flickering
on early September and late October observations does not necessarily imply
it is always absent. Unfortunately, at the time of flickering observations,
the object was no more observable from Italy, and therefore we have no
contemporaneous spectroscopy to compare the absence of flickering to the
amount of emission in H$\alpha$.

	\begin{figure}
	\centerline{
		\includegraphics[width=12.cm,clip=]{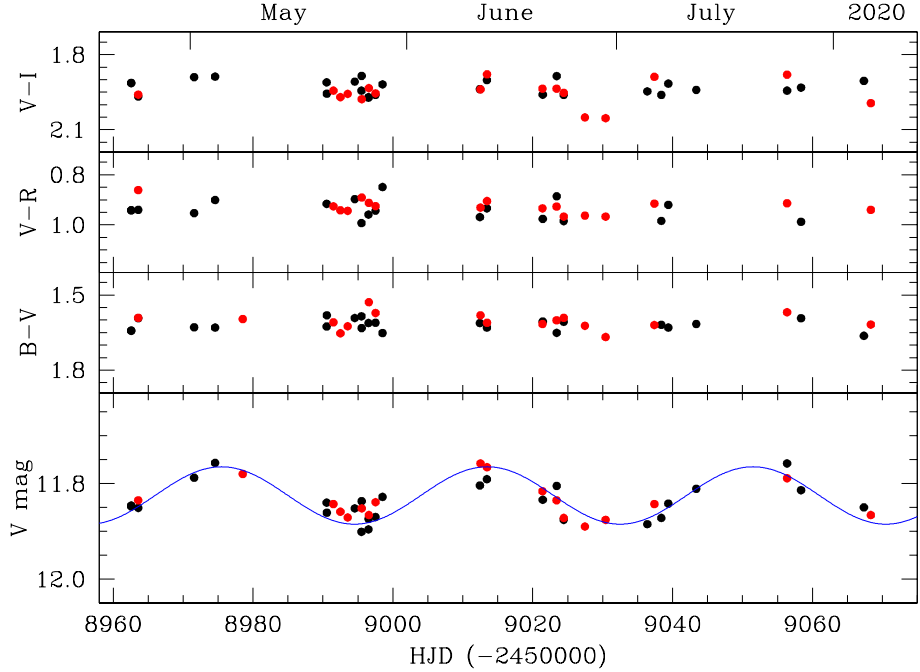}}
	\caption{Our April-to-August 2020 light-curve of 2SXPS
        J173508.4-292958 built from data in Table~1 (red dots mark
        the observations obtained with ANS Collaboration telescope ID~0310,
        and black dots those with telescope ID 1301).  The sinusoid going
        through the $V$-band data has minima according to Ephemeris (1), an
        amplitude of 0.12 mag and a period of 38 days.}
	\label{f5}
	\end{figure}

\subsection{Photometric evolution}

Our $B$$V$$R$$I$ photometry from Table~1 is plotted in Fig.~\ref{f5}.  The $V$
band light-curve is characterized by a mean 11.825 mag value, and a
sinusoidal modulation with a period of 38 days and an amplitude of 0.12
mag, with the following ephemeris providing times of mimina:
\begin{equation}
{\rm min(V)} = 2458994.5(\pm 1) + 38(\pm 0.5)\times E
\end{equation}
The colors do not appear to change along the cycle.  As illustrated by
the SED in Fig.~\ref{f1}, the emission at optical wavelengths is dominated by the
K4III giant, so the sinusoidal variability probably originates from it,
either in the form of some pulsation or the rotation of a spotted surface.

The sinusoidal variability could instead point to an ellipsoidal deformation
of the K4III, should the latter fill its Roche lobe, in which case the orbital
period would be twice the photometric one, or 76 days.  The typical mass of
a field K4III is listed as 1.2~M$_\odot$ by Drilling and Landolt (2000).
Assuming 1.0~M$_\odot$ for the companion, this leads to a semi-major axis
for the binary system of 98~R$_\odot$.  The relation between mass ratio $q$,
orbital separation $a$, and Roche lobe radius $R_{RL}$ as given by 
Eggleton (1983) is
\begin{equation}
\frac{R_{RL}}{a} = \frac{0.49 q^{2/3}}{0.6q^{2/3} + \ln(1 + q^{1/3})}
\end{equation}
and returns a radius of 39~R$_\odot$ for the K4III, twice the tabulated
values of $\sim$20~R$_\odot$ (Drilling and Landolt 2000). The discrepancy
would persist also for different assumption on the mass of the companion:
0.5~M$_\odot$ leads to 41~R$_\odot$ and 2.5~M$_\odot$ to 37~R$_\odot$.
Therefore, the origin of the observed sinusoidal light-curve is probably
other that Roche-lobe filling.

	\begin{figure}
	\centerline{
		\includegraphics[width=12.cm,clip=]{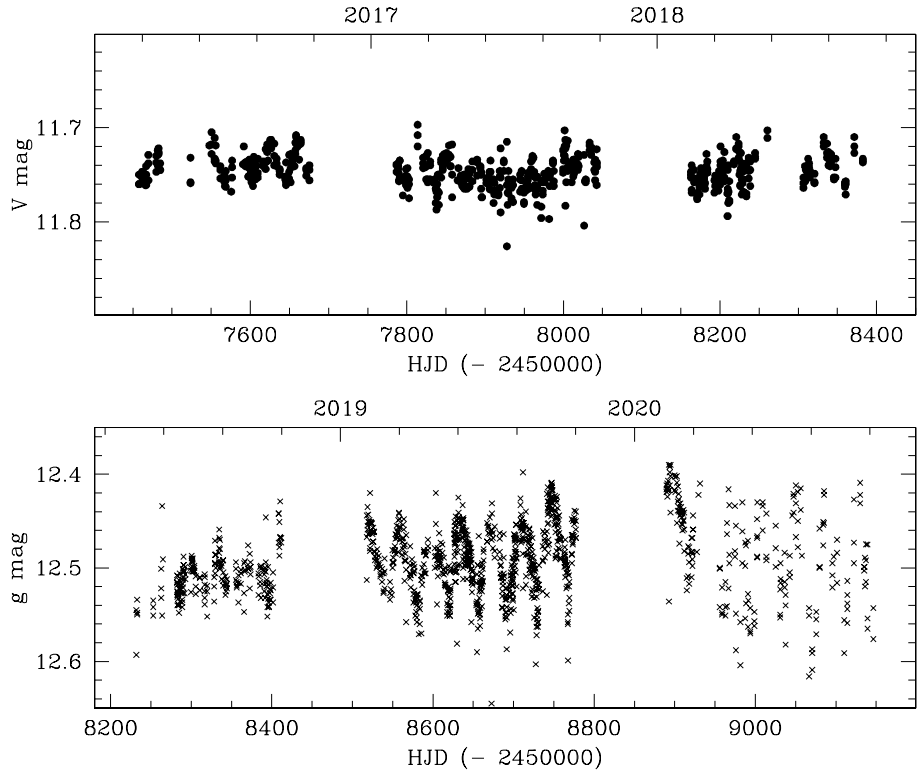}}
	\caption{Long-term light-curve of 2SXPS J173508.4-292958 built from
         ASAS-SN sky-patrol data in $V$ and $g$ bands.}
	\label{f6}
	\end{figure}

	\begin{figure}
	\centerline{
		\includegraphics[width=12.cm,clip=]{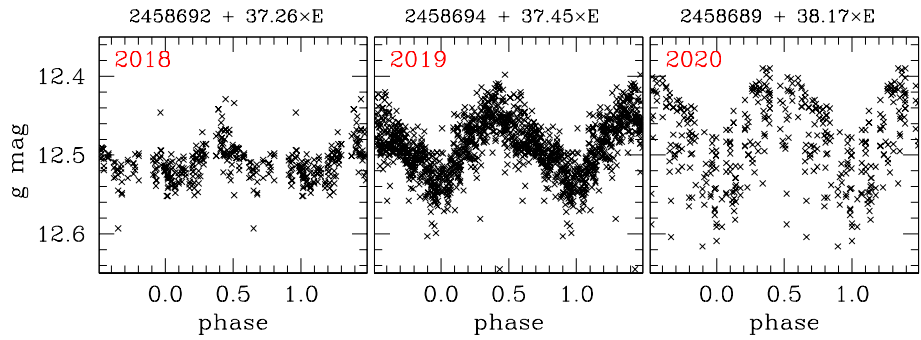}}
	\caption{Phased light-curves for the modulated part of Figure~6.}
	\label{f7}
	\end{figure}

In support of this conclusion comes the absence of detectable radial
velocity changes in Table~3.  A 76-day period would correspond to an orbital
velocity of the K4III of 30~km~s$^{-1}$ for a 0.5~M$_\odot$ companion and
38~km~s$^{-1}$ for 2.5~M$_\odot$.  The maximum photometric amplitude (for
edge-on systems) of the ellipsoidal modulation is 0.25/0.30 mag.  An
amplitude of 0.12 mag for 2SXPS requires a significant orbital inclination,
such that the amplitude of radial velocity variations would be at least 10
~km~s$^{-1}$, incompatible with results from Table~3.

The nail in the coffin about ellipsoidal modulation is however the fact that
the sinusoidal variability is not always present but highly variable with
time, as discussed in the following section.

\subsection{Long-term light-curve}

To reconstruct a longer term light-curve of 2SXPS we have downloaded ASAS-SN
all-sky patrol data (Shappee et al.  2014, Kochanek et al.  2017) and
plotted them in Fig.~\ref{f6}. 2SXPS was observed by ASAS-SN with a
$V$-filter over 2016-2018 and with a $g$-filter during 2018-2020.

The median value of the 2016-2018 ASAS-SN $V$ band data in Fugure~6 is
0.076~mag brighter than our data in 2020.  We do not regard this as
significative: ASAS-SN data are just differential magnitudes with respect to
{\it not-transformed} data for field stars, while our photometry is fully
transformed to Landolt's system of equatorial standards (cf.  sect.~2.1). 
The rather red color of 2SXPS, much redder than typical field stars, can
easily account for the offset affecting ASAS-SN data.

The ASAS-SN light-curve in Fig.~\ref{f6} shows a stable 2SXPS during 2016-2018,
and appearance of the sinusoidal modulation only in 2019, which extended to
2020 with a lower degree of coherence.  The phased light-curves of 2SXPS from
ASAS-SN data for 2018, 2019 and 2020 is presented in Fig.~\ref{f7}, with the
best-fit ephemeris given at the top of the panels. The differences
between them are marginally within uncertainties.

While certainly arguing against an interpretation in terms of ellipsoidal
distortion of the K4III giant, the light-curves of Fig.~\ref{f5}, \ref{f6}
and \ref{f7} do not disentangle a specific cause for the variability
observed in 2SXPS, although some kind of pulsation could be a viable
explanation.

\section{Conclusions}

Our observations suggest 2SXPS to be a previously unknown symbiotic star, of
the accreting-only variety (AO-SySt).  The low space velocity, absence of
cool circumstellar dust, solar-like metallicity and lack of enhancement in
Barium set 2SXPS aside from other symbiotic stars of the {\it yellow} class
(i.e.  those harbouring a cool giant of the F,G,K spectral types).  As
typical of AO-SySt systems, the signature for 2SXPS of on-going accretion
onto a degenerate companion are feeble (weak and structure emission in
H$\alpha$, just a hint of near-UV excess emission) and highly variable in
time, so that the AO-SySt nature may be not recognized at all times from
optical observations alone.  The star clearly deserves further
observations, especially accurate radial velocities over a long time
interval in order to derive the orbital period, orbital inclination and an
estimate for the mass of the unseen companion.

\end{document}